\documentstyle[a4,12pt]{article}
\def\U4s{\Upsilon (4S)}
\def\ra{\rightarrow}
\def\be{\begin{equation}}
\def\ee{\end{equation}}
\def\ba{\begin{eqnarray}}
\def\ea{\end{eqnarray}}

\begin{document}
\parskip=5pt plus 1pt minus 1pt

\begin{flushright}
{\small\bf DPNU-97-16} \\
{March 1997}
\end{flushright}

\vspace{0.2cm}

\begin{center}
{\large\bf $CP$-violating Signals in the First-round Experiments} \\
{\large\bf of a $B$-meson Factory}
\footnote{Invited talk presented at the 1997 Shizuoka Workshop on Masses and
Mixings of Quarks and Leptons, Shizuoka, March 19 - 21 (to appear in
the workshop proceedings).}
\end{center}

\vspace{0.3cm}

\begin{center}
{\bf Zhi-zhong Xing} \footnote{Electronic address: xing@eken.phys.nagoya-u.ac.jp}
\end{center}
\begin{center}
{\it Department of Physics, Nagoya University, Chikusa-ku, Nagoya 464-01, Japan}
\end{center}

\vspace{2.5cm}

\begin{abstract}
Determination of the characteristic $CP$-violating quantity $\sin (2\beta)$ should be the 
central goal of a $B$-meson factory in its first-round experiments. Except the
gold-plated channels $B_d\rightarrow \psi K_S$ and $\psi K_L$, three other types
of $B_d$ decays to $CP$ eigenstates can also serve for the extraction of $\sin (2\beta)$
in the standard model: (a) the $CP$-forbidden transitions $(B_d\bar{B}_d)_{\Upsilon (4S)} 
\rightarrow (X_c K_S) (X_c K_S)$ and $(X_c K_L) (X_c K_L)$, where $X_c = \psi, 
\psi^{\prime},\eta_c$, etc; (b) the decay modes $B_d\rightarrow D^{(*)+}D^{(*)-}$ 
and $D^{(*)0}\bar{D}^{(*)0}$, whose amplitudes have simple isospin relations; and
(c) the decay modes $B_d\rightarrow (f_{CP})_D + (\pi^0, \rho^0, a^0_1, ~ {\rm etc})$,
in which $f_{CP}$ is a $CP$ eigenstate (such as $\pi^+\pi^-$, $K^+K^-$ or $K_S \pi^0$)
arising from either $D^0$ or $\bar{D}^0$ in the neglect of $D^0$-$\bar{D}^0$ mixing.
We carry out an analysis of the $CP$-violating signals existing in these typical processes, 
without loss of the possibility that new physics might significantly affect 
$B^0_d$-$\bar{B}^0_d$ or $K^0$-$\bar{K}^0$ mixing.
We also show that the magnitude of $\sin (2\beta)$ can be well determined, in terms of
only $|V_{us}|$, $m_d/m_s$ and $m_u/m_c$, from a variety of quark mass ans$\rm\ddot{a}$tze.
\end{abstract}

\newpage

%===========================================================================
\section{Introduction}

The origin of $CP$-violating phenomena, observed in neutral kaon decays, has been an
intriguing puzzle of particle physics. Among various proposed 
mechanisms of $CP$ violation \cite{CPbooks}, the most natural and economical one is the
Kobayashi-Maskawa (KM) picture which works within the standard electroweak
model \cite{KM}. It is expected that large and theoretically clean signals of $CP$
violation, induced purely by the nontrivial phase of the KM matrix, may
manifest themselves in some neutral $B$-meson decays to $CP$ eigenstates
\cite{Sanda80,BigiSanda81}. This possibility has attracted a lot of phenomenological
interest \cite{Quinn96}, leading experimentally to the $B$ factory programs at KEK, SLAC,
DESY and LHC (as well as the upgrades of the existing facilities at
Cornell and Fermilab).

The central goal of the first-round experiments at a $B$-meson factory
should be to determine the $CP$-violating phase
\be
\beta \; \equiv \; \arg \left ( - ~ \frac{V^*_{tb}V_{td}}{V^*_{cb}V_{cd}}
\right ) \; ,
%		(1.1)
\ee
which represents one angle of the KM unitarity triangle
$V^*_{ub}V_{ud} + V^*_{cb}V_{cd} + V^*_{tb}V_{td} = 0$ in the complex plane.
The standard model predicts $CP$ asymmetries of the magnitude $\sin (2\beta)$,
arising from the interference of decay and $B^0_d$-$\bar{B}^0_d$ mixing, in
some $B_d$ decay modes such as $B^0_d$ vs $\bar{B}^0_d\rightarrow \psi K_S$
and $\psi K_L$. The number of $B^0_d\bar{B}^0_d$ events needed for the
pragmatic measurement of $\sin (2\beta)$ to three standard deviations can be
estimated as follows:
\be
{\cal N}_{B\bar{B}} \; = \; \left [ \frac{3}{\sin (2\beta)} \right ]^2
\frac{1}{{\cal B}_{\rm eff} ~ \epsilon^{~}_{\rm com}} \; , 
%		(1.2)
\ee
where ${\cal B}_{\rm eff}$ is the effective branching fraction of $B^0_d$
or $\bar{B}^0_d$ decaying to a $CP$ eigenstate, and $\epsilon^{~}_{\rm com}$
is the composite detection efficiency of the decay mode under consideration.
An analysis of current experimental data on $|V_{ub}/V_{cb}|$, $B^0_d$-$\bar{B}^0_d$
mixing and $\epsilon^{~}_K$ yields the constraint $0.32\leq \sin (2\beta) \leq 
0.94$ \cite{AliLondon96}. If we assume $\epsilon^{~}_{\rm com} = 10\%$ and
${\cal N}_{B\bar{B}} = 10^7$ (or $10^8$) in the first-round experiments of a
$B$ factory, then the size of ${\cal B}_{\rm eff}$ is required to
be $3.6\times 10^{-5}$ (or $3.6\times 10^{-6}$) for $\sin (2\beta) =0.5$.
Taking into account the fact that ${\cal B}_{\rm eff}$ should include the
cost for flavor tagging of the parent $B^0_d$ and $\bar{B}^0_d$ mesons, 
one has to choose those $B_d$ decays of interest whose branching ratios are as large
as possible.

The gold-plated decay modes for the extraction of $\sin (2\beta)$ are expected to be
$B_d\rightarrow \psi K_S$ and $B_d\rightarrow \psi K_L$ \cite{Sanda80,BigiSanda81}. 
Their decay amplitudes are governed by 
\be
a_2 ~ |V_{cb} V_{cs}| \; \approx \; A \lambda^2 a_2 \; \sim \; 8.9 \times 10^{-3} \; 
%		(1.3)
\ee
in the naive factorization approximation, where $A$ ($\approx 0.8$) and $\lambda$
($\approx 0.22$) are the Wolfenstein parameters \cite{Wolfenstein83} 
and $a_2$ ($\approx 0.23$) is the Bauer-Stech-Wirbel (BSW) factorization 
coefficient \cite{BSW,Browder96}. There are other two types of 
$B_d$ decays to $CP$ eigenstates, which have branching ratios comparable in
magnitude with that of $B^0_d\rightarrow \psi K_S$. One typical example is 
$B_d\rightarrow D^+D^-$, whose decay amplitude is dominated by
\be
a_1 ~ |V_{cb} V_{cd}| \; \approx \; A \lambda^3 a_1 \; \sim \; 8.8 \times 10^{-3} \; ,
%		(1.4)
\ee
where $a_1$ ($\approx 1.03$) is the other BSW factorization coefficient \cite{Browder96}. 
Another typical example is $B_d \rightarrow (f_{CP})_D + \pi^0$, in which $f_{CP}$ 
is a $CP$ eigenstate coming from $D^0$ or $\bar{D}^0$ in the neglect of $D^0$-$\bar{D}^0$
mixing \cite{DunietzSnyder91}. The primary transition amplitude of this decay mode is associated 
dominantly with 
\be
a_2 ~ |V_{cb} V_{ud}| \; \approx \; A \lambda^2 a_2 \; \sim \; 8.9 \times 10^{-3} \; .
%		(1.5)
\ee
$CP$ asymmetries in all three types of decays mentioned above are dominated by
$\sin (2\beta)$ within the standard model.

In this talk we shall present a three-plus-one strategy to determine
the $CP$-violating observable $\sin (2\beta)$. Starting from a variety of quark
mass ans$\rm\ddot{a}$tze, we can calculate the KM matrix in terms of quark mass
ratios and a $CP$-violating phase. The magnitude of $\sin (2\beta)$
is predictable, as shown in section 2, by use of the well-determined quantities
$|V_{us}|$, $m_d/m_s$ and $m_u/m_c$. Nontrivially, section 3 is devoted to 
$CP$-forbidden decays of the type $(B_d\bar{B}_d)_{\Upsilon (4S)}\rightarrow
(\psi K_S) (\psi K_S)$ or $(\psi K_L) (\psi K_L)$, whose decay rates are 
proportional to $\sin^2(2\beta)$ in the standard model. In section 4, we carry out
an isospin analysis of $CP$ violation in $B_d\rightarrow D^+D^-$ and $D^0\bar{D}^0$
to extract $\sin (2\beta)$ and probe the penguin-induced phase information.
The possibility to determine $\sin (2\beta)$ in decay modes of the type
$B_d \rightarrow (f_{CP})_D + \pi^0$ is discussed in section 5. A brief summary
of our main results, together with some further discussions, is included in section 6.

It is worthwhile to point out that the explicit analyses in the subsequent
sections allow the presence of new physics in $B^0_d$-$\bar{B}^0_d$ mixing and
$K^0$-$\bar{K}^0$ mixing. The relevant $CP$-violating signals turn out to be
$\sin (2\beta)$ if we adopt the standard model predictions for the mixing
phases, i.e.,
\be
\frac{q^{~}_B}{p^{~}_B} \; = \; \frac{V^*_{tb}V_{td}}{V_{tb}V^*_{td}} \; ,
~~~~~~~~
\frac{q^{~}_K}{p^{~}_K} \; = \; \frac{V^*_{cs}V_{cd}}{V_{cs}V^*_{cd}} \; .
%		(1.6)
\ee
Thus most of our
results are also valid beyond the standard model, and they should be useful 
for the experimental studies to be carried out at the forthcoming $B$-meson
factories.

%============================================================================
\section{Determination of $\sin (2\beta)$ from mass ans$\bf\ddot{a}$tze}
\setcounter{equation}{0}

It is expected that flavor mixing parameters can be completely predicted 
from fermion mass matrices in the framework of a theory beyond the 
standard model. Before the success in finding this more fundamental theory,
the phenomenological approach is to look for the most proper pattern of
quark mass matrices which are able to result in the experimentally favored
relations between KM matrix elements and quark mass ratios 
\cite{Fritzsch78,GeorgiJarlskog}. The relevant
symmetries hidden in such quark mass ans$\rm\ddot{a}$tze may provide 
useful hints towards the dynamical details of fermion mass generation and
$CP$ violation \cite{Fritzsch97}. 

Here let us illustrate a variety of quark mass ans$\rm\ddot{a}$tze in order 
to predict the magnitude of $\sin (2\beta)$. In
the standard model or its extensions which have no flavor-changing
right-handed currents, we can choose the up and down quark mass matrices 
(denoted by $M_{\rm u}$ and $M_{\rm d}$, respectively) to be Hermitian
without loss of generality \cite{Jarlskog85}.  
We also assume that $M_{\rm u}$ and $M_{\rm d}$
have the parallel structures (i.e., parallel hierarchies and texture zeros),
coming naturally from the same dynamics. After the diagonalization of 
$M_{\rm u,d}$ through the unitary transformation $O^{\dagger}_{\rm u,d}M_{\rm u,d}
O_{\rm u,d}$, one obtains the mass eigenvalues. The KM matrix in the charged 
weak currents turns out to be $V\equiv O^{\dagger}_{\rm u} O_{\rm d}$. 
Taking into account the facts 
\ba
m_u & \ll & m_c \; \ll \; m_t \; , \nonumber \\
m_d & \ll & m_s \; \ll \; m_b \; ,
%		(2.1)
\ea
and \cite{XingV}
\ba
|V_{tb}| \; > \; |V_{ud}| \; > \; |V_{cs}| & \gg & |V_{us}| \; > \; |V_{cd}| \; \nonumber \\
& \gg & |V_{cb}| \; > \; |V_{ts}| \; \nonumber \\
& \gg & |V_{td}| \; > \; |V_{ub}| \; > \; 0 \; ,
%		(2.2)
\ea
we can draw the following points:

(a) $M^{\rm u}_{11} = M^{\rm d}_{11} =0$ (or $|M^{\rm u,d}_{11}| \ll |M^{\rm u,d}_{12}|$)
is a sufficient condition to get proper $|V_{us}|$ and $|V_{cd}|$ in leading order
approximations:
\ba
V_{us} & \approx & \sqrt{\frac{m_d}{m_s}} \; ~ - ~ \; \exp ({\rm i} \varphi^{~}_{12}) ~ \sqrt{\frac{m_u}{m_c}} \; ,
\nonumber \\
V_{cd} & \approx & \sqrt{\frac{m_u}{m_c}} \; ~ - ~ \; \exp ({\rm i} \varphi^{~}_{12}) ~ \sqrt{\frac{m_d}{m_s}} \; ,
%		(2.3)
\ea
where $\varphi^{~}_{12} \equiv \arg (M^{\rm u}_{12} / M^{\rm d}_{12})$ is a phase
parameter. These two relations form two congruent triangles in the complex plane, the so-called Cabibbo
triangles \cite{FritzschXing95}. 

(b) $M^{\rm u}_{13} = M^{\rm d}_{13} =0$ (or $|M^{\rm u,d}_{13}| \ll |M^{\rm u,d}_{23}|$),
together with condition (a), can approximately lead to 
\be
\left | \frac{V_{ub}}{V_{cb}} \right | \; \approx \; \sqrt{\frac{m_u}{m_c}} \; , ~~~~~~~~
\left | \frac{V_{td}}{V_{ts}} \right | \; \approx \; \sqrt{\frac{m_d}{m_s}} \; .
%		(2.4)
\ee
We observe that $|V_{ub}/V_{cb}| < |V_{td}/V_{ts}|$ due to the fact $m_u/m_c < m_d/m_s$.
It is worth mentioning that relations (2.3) and (2.4) are basically the results of the
Fritzsch ansatz, which has texture zeros $M^{\rm u,d}_{11} = M^{\rm u,d}_{22} = M^{\rm u,d}_{13} =0$
\cite{Fritzsch78}.

(c) $|V_{cb}|$ (or $|V_{ts}|$) depends upon the relative size between $M^{\rm u,d}_{22}$
and $M^{\rm u,d}_{23}$. If they are comparable in magnitude, we arrive at 
\be
|V_{cb}| \; \approx \; |V_{ts}| \; \approx \; \left | \left | \frac{M^{\rm d}_{23}}{M^{\rm d}_{22}}
\right |  \frac{m_s}{m_b} ~ - ~ \exp ({\rm i} \varphi^{~}_{23})  \left | 
\frac{M^{\rm u}_{23}}{M^{\rm u}_{22}} \right | \frac{m_c}{m_t} \right | \; 
%		(2.5)
\ee
in leading order approximations, where $\varphi^{~}_{23} \equiv \arg (M^{\rm u}_{23}/M^{\rm d}_{23})$.
It can be shown that the contribution of $\varphi^{~}_{23}$ to $CP$ violation in the KM matrix
is negligibly small. 

Indeed conditions (a) and (b) imply that the Hermitian mass matrices $M_{\rm u}$ and $M_{\rm d}$
should take the following generic form:
\be
\left ( \matrix{
0	& \times	& 0 \cr
\times^*	& \bigtriangleup	& \bigtriangledown \cr
0	& \bigtriangledown^*	& \Box \cr} \right ) \; .
%		(2.6)
\ee
The key point of the above quark mass ans$\rm\ddot{a}$tze is that either of the two Cabibbo
triangles can be rescaled by $V^*_{cb}$ or $V^*_{ts}$, and the resultant triangle is congruent
approximately with the unitarity triangle $V^*_{ub}V_{ud} + V^*_{cb}V_{cd} + V^*_{tb}V_{td}=0$
\cite{Fritzsch97,FritzschXing95}. Then three angles of the unitarity triangle
are determinable from three sides of the Cabibbo triangle, which are nearly independent
of the mass ratios $m_c/m_t$ and $m_s/m_b$ as well as the phase parameter $\varphi^{~}_{23}$.
For our present purpose, we only write out the expression of $\sin (2\beta)$:
\small
\be
\sin (2\beta) \; \approx \; \frac{1}{2} \left [ \frac{m_s}{m_d} + \frac{1}{|V_{us}|^2}
\left ( 1 - \frac{m_u}{m_c} \cdot \frac{m_s}{m_d} \right ) \right ]
\sqrt{ 4\frac{m_u}{m_c} \cdot \frac{m_d}{m_s} - \left ( \frac{m_u}{m_c} + \frac{m_d}{m_s}
- |V_{us}|^2 \right )^2 } \; .
%		(2.7)
\ee
\normalsize
So far $|V_{us}|$ has been precisely measured \cite{PDG}: $|V_{us}| = 0.2205 \pm 0.0018$.
The latest result of the chiral perturbation theory yields $m_s/m_d = 19.3\pm 0.9$
and $m_u= 5.1 \pm 0.9$ MeV at the scale 1 GeV \cite{Leutwyler96}. The value of $m_c$(1 GeV)
is expected to be in the range 1.0 $-$ 1.6 GeV, or around 1.35 GeV \cite{PDG,Leutwyler82}.
With these inputs, we calculate $\sin (2\beta)$ and plot the result in Fig. 1. One can see
that the prediction of quark mass ans$\rm\ddot{a}$tze for $\sin (2\beta)$ is quite restrictive
in spite of some errors associated with quark masses. It lies in the experimentally
allowed region $0.32 \leq \sin (2\beta) \leq 0.94$, obtained from the analysis of current 
data on $|V_{ub}/V_{cb}|$, $B^0_d$-$\bar{B}^0_d$ mixing and $\epsilon^{~}_K$ within the 
standard model \cite{AliLondon96}.

In the above discussions, we did not assume any specific theory (or model) that can 
naturally guarantee conditions (a) and (b) for $M_{\rm u}$ and $M_{\rm d}$. It is very possible
that such a theory exists at a superheavy energy scale (e.g., the scale of string
theories or that of grand unification theories). Fortunately, the instructive relations (2.3) and
(2.4) are independent of the renormalization-group effects to a good degree of accuracy
\cite{XingM}; 
in other words, they hold at both very high and very low energy scales. Thus the prediction
(2.7) remains valid even if $M_{\rm u}$ and $M_{\rm d}$ are derived at a superheavy scale,
and it can be confronted directly with the low-energy experimental data. In contrast, relation
(2.5) will be spoiled by the renormalization-group effects, since $|V_{cb}|$ (or $|V_{ts}|$),
$m_s/m_b$ and $m_c/m_t$ may have quite different evolution behaviors with energy scales
(see, e.g., \cite{XingM}).

%========================================================================================
\section{Probing $\sin (2\beta)$ in $(B_d\bar{B}_d)_{\U4s} \ra 
(\psi K_S) (\psi K_S)$}
\setcounter{equation}{0}

It was first pointed out by Wolfenstein \cite{Wolfenstein84} 
that the search for $CP$-forbidden transitions of the type 
\be
\left ( B_d \bar{B}_d \right )_{\U4s} \; \longrightarrow \; \left (f_a f_b \right )_{CP\rm -even} \; ,
%		(3.1)
\ee
where $f_a$ and $f_b$ denote two $CP$ eigenstates with the same $CP$ parity (in contrast,
the initial state has the $CP$-odd parity), would serve as a distinctive test of $CP$
violation in the $B^0_d$-$\bar{B}^0_d$ system. For such a joint decay mode, the $CP$-violating
signal can be established by measuring the decay rate itself other than the decay rate
asymmetry. In practical experiments, this implies that neither flavor tagging (for the
parent $B_d$ mesons) nor time-dependent measurements (for the whole decay chain) are
necessary. The feasibility of detecting reaction (3.1) depends mainly upon the 
branching ratios ${\cal B}(B_d^0\ra f_a)$ and ${\cal B}(B^0_d\ra f_b)$. 

The most interesting $CP$-forbidden channels on the $\U4s$ resonance should be 
\ba
\left (B_d \bar{B}_d \right )_{\U4s} & \longrightarrow & \left (X_c K_S \right )_{B_d}
\left (X_c K_S \right )_{\bar{B}_d} \; , \nonumber \\
\left (B_d \bar{B}_d \right )_{\U4s} & \longrightarrow & \left (X_c K_L \right )_{B_d}
\left (X_c K_L \right )_{\bar{B}_d} \; , 
%		(3.2)
\ea
in which $X_c$ stands for a set of possible charmonium states that can form $CP$
eigenstates with $K_S$ ($CP$-odd or $CP$-even) and $K_L$ ($CP$-even or $CP$-odd).
The typical examples may include 
\footnote{Note that $\psi^{\prime}\ra \psi \pi\pi$, $\psi^{\prime\prime}\ra D\bar{D}$,
and $\eta^{\prime}_c \ra \eta_c \pi\pi$.} 
\be
X_c \; = \; \psi \; , ~ \psi^{\prime} \; , ~ \psi^{\prime\prime} \; , ~
\eta_c \; , ~ \eta^{\prime}_c \; , ~ {\rm etc.}
%		(3.3)
\ee
Since all the transitions $B_d^0\ra X_c K_S$ occur through the same weak interactions,
their branching ratios should be comparable in magnitude. Neglecting tiny $CP$ violation
in the kaon system, we have ${\cal B} (B^0_d\ra X_c K_S) = {\cal B} (B^0_d\ra X_c K_L)$
to an excellent degree of accuracy. In contrast with (3.2), the transitions
\ba
\left (B_d \bar{B}_d \right )_{\U4s} & \longrightarrow & \left (X_c K_S \right )_{B_d}
\left (X_c K_L \right )_{\bar{B}_d} \; , \nonumber \\
\left (B_d \bar{B}_d \right )_{\U4s} & \longrightarrow & \left (X_c K_L \right )_{B_d}
\left (X_c K_S \right )_{\bar{B}_d} \; 
%		(3.4)
\ea
are allowed by $CP$ symmetry. If we make use of ${\cal R} (K_S, K_S)$, ${\cal R} (K_L,
K_L)$, ${\cal R} (K_S, K_L)$ and ${\cal R} (K_L, K_S)$ to respectively denote the rates of 
the above four types of joint decays, then 
\be
{\cal S}_{CP} \; \equiv \; \frac{ {\cal R} (K_S, K_S) ~ + ~ {\cal R} (K_L, K_L)}
{ {\cal R} (K_S, K_S) + {\cal R} (K_S, K_L) + {\cal R} (K_L, K_S) 
+ {\cal R} (K_L, K_L) } \; 
%		(3.5)
\ee
is a clean signal of $CP$ violation independent of the ambiguity from hadronic matrix
elements. Furthermore, a sum over all possible $X_c$ states as listed in (3.3) can
enhance the statistics of a single mode (say, $X_c = \psi$) by several times (even
one order) \cite{XingB96}, without
dilution of the $CP$-violating signal ${\cal S}_{CP}$. Only if the combined branching
fraction of $(B_d\bar{B}_d)_{\U4s} \ra (X_c K_S) (X_c K_S)$ can amount to 
$10^{-6}$ or so, a signal of the magnitude ${\cal S}_{CP}\sim 0.1$ should be explored in the 
first-round experiments of an $e^+e^-$ $B$-meson factory.

The generic formulas for coherent $B_d\bar{B}_d$ decays have been presented in
the literature (see, e.g., \cite{BigiSanda81,XingB96,BigiSanda88}). 
Explicitly, the time-independent decay rate of 
$(B_d \bar{B}_d)_{\U4s} \rightarrow (f_a f_b)$ can be written as
\ba
{\cal R} (f_a, f_b) & = & N_f \left \{ \frac{x^2_d}{1+x^2_d} \left [ 1 + |\rho^{~}_a|^2
|\rho^{~}_b|^2  - 2 {\rm Re} \left ( \frac{q^{~}_B p^*_B}{p^{~}_B q^*_B} 
\rho^{~}_a \rho^{~}_b \right ) \right ] \right . \nonumber \\
&  & \left . + ~ \frac{2+x^2_d}{1+x^2_d} \left [ |\rho^{~}_a|^2
+ |\rho^{~}_b|^2 - 2 {\rm Re} \left ( \rho^{~}_a \rho^{~}_b \right ) \right ]
\right \} \; ,
%		(3.6)
\ea
where $N_f$ is a normalization factor proportional to the product of the
decay rates of $B_d^0\ra f_a$ and $B_d^0\ra f_b$; $x_d\equiv \Delta m/\Gamma 
\approx 0.73$ is the $B^0_d$-$\bar{B}^0_d$ mixing parameter \cite{PDG}; $q^{~}_B/p^{~}_B$
signifies the phase information from $B^0_d$-$\bar{B}^0_d$ mixing; and
$\rho^{~}_{a,b}$ are ratios of the decay amplitudes $A (\bar{B}^0_d\ra
f_{a,b})$ to $A (B^0_d\ra f_{a,b})$. In obtaining the above formula, we have
neglected the tiny $CP$ violation induced purely by $B^0_d$-$\bar{B}^0_d$ mixing
(i.e., $|q^{~}_B/p^{~}_B|=1$ is taken). Within the standard model, 
$q^{~}_B/p^{~}_B = (V^*_{tb}V_{td})/(V_{tb}V^*_{td})$ given in (1.6) 
is a good approximation.

For the cases of $f_{a,b} = X_c K_{S,L}$, $\rho^{~}_{a,b}$ turn out to be
\be
\rho^{~}_{X_c K_S} \; = \; - \rho^{~}_{X_c K_L} \; = \; 
\pm  ~ \frac{q^*_K}{p^*_K} ~ \frac{V_{cb} V_{cs}^*}{V^*_{cb}V_{cs}} \; ,
%		(3.7)
\ee
where ``$\pm$'' is the $CP$ parity of $|X_c K_S\rangle$ state, and
$q^{~}_K/p^{~}_K$ stands for the phase information from $K^0$-$\bar{K}^0$
mixing in the final state (here $|q^{~}_K/p^{~}_K|=1$ is assumed). Defining
the phase parameter
\be
\phi_{\psi K} \; \equiv \; \frac{1}{2} \arg \left ( \frac{q^{~}_B}{p^{~}_B} ~
\frac{q^*_K}{p^*_K} ~ \frac{V_{cb} V^*_{cs}}{V^*_{cb} V_{cs}} \right ) \; ,
%		(3.8)
\ee
then we obtain the following decay rates:
\ba
{\cal R} (K_S, K_S ) & = & {\cal R} (K_L, K_L) \; = \; 4 N_{X_cK} \left [
\frac{x^2_d}{1+x^2_d} \sin^2 (2\phi_{\psi K}) \right ] \; , \nonumber \\
{\cal R} (K_S, K_L ) & = & {\cal R} (K_L, K_S) \; = \; 4 N_{X_cK} \left [
2 ~ - ~ \frac{x^2_d}{1+x^2_d} \sin^2 (2\phi_{\psi K}) \right ] \; , 
%		(3.9)
\ea
where the normalization factor $N_{X_c K}$ is proportional to the square of
the decay rate of $B_d^0\rightarrow X_c K_S$. As a result, the $CP$-violating signal
${\cal S}_{CP}$ defined in (3.5) reads 
\be
{\cal S}_{CP} \; = \; \frac{1}{2} \frac{x^2_d}{1+x^2_d} \sin^2 (2\phi_{\psi K} ) \; ,
%		(3.10)
\ee
purely determined by the $B^0_d$-$\bar{B}^0_d$ mixing parameter $x_d$ and the
combined weak phase $\phi_{\psi K}$. In the standard model, we get $\phi_{\psi K} = \beta$
to a good degree of accuracy.
If new physics affects $B^0_d$-$\bar{B}^0_d$ mixing and (or) $K^0$-$\bar{K}^0$
mixing, however, $\phi_{\psi K}$ could significantly deviate from $\beta$.

For illustration, we plot the magnitude of ${\cal S}_{CP}$ as a function of
$\phi_{\psi K}$ in Fig. 2, with the input $x_d \approx 0.73$. Current constraint 
on $\beta$ is $9.3^{\circ} \leq \beta \leq 35^{\circ}$ \cite{AliLondon96}, at $95\%$ confidence
level in the standard model. We see that there is large room for $\phi_{\psi K}$
or ${\cal S}_{CP}$ to accommodate new physics. The maximal value of ${\cal S}_{CP}$
($\approx 0.17$) can be obtained when $\phi_{\psi K} = \pm 90^{\circ}$. 

%=====================================================================
\section{Probing $\sin (2\beta)$ in $B_d \rightarrow D^+D^-$ and $D^0\bar{D}^0$}
\setcounter{equation}{0}

The measurement of $CP$ asymmetries in $B_d\rightarrow 
D^+D^-$ and $D^0\bar{D}^0$ can not only cross-check the extraction of $\beta$ from
decays of the type $B_d\rightarrow \psi K_S$, but also shed some light on the penguin
effects and final-state interactions in nonleptonic $B$
decays to double charmed mesons. For this reason, it is worth studying
$B_d\rightarrow D^+D^-$ and $D^0\bar{D}^0$ 
in a model-independent approach. The similar treatment is applicable
to the processes $B_d\rightarrow D\bar{D}^*$, $D^*\bar{D}$, etc.

Let us carry out an isospin analysis of the decay modes $B\rightarrow D\bar{D}$,
to relate their weak and strong phases to the relevant observables \cite{SandaXing}.
The effective weak Hamiltonians responsible for 
$B^-_u\rightarrow D^-D^0$, $\bar{B}^0_d\rightarrow D^+D^-$,
$\bar{B}^0_d\rightarrow D^0\bar{D}^0$ and their $CP$-conjugate 
processes have the isospin structures $|1/2, -1/2\rangle$
and $|1/2, +1/2\rangle$ respectively. The decay amplitudes of these 
transitions can be written in terms of the $I=1$ and $I=0$ isospin amplitudes:
\begin{eqnarray}
A^{+-} & \equiv & \langle D^+D^- |{\cal H}_{\rm eff}| B^0_d\rangle \; = \;
\frac{1}{2} \left ( A_1 ~ + ~ A_0 \right ) \; , \nonumber \\
A^{00} & \equiv & \langle D^0\bar{D}^0 |{\cal H}_{\rm eff}| B^0_d\rangle \; = \;
\frac{1}{2} \left ( A_1 ~ - ~ A_0 \right ) \; , \nonumber \\
A^{+0} & \equiv & \langle D^+\bar{D}^0 |{\cal H}_{\rm eff}| B^+_u\rangle \; = \; A_1 \; ;
%		(4.1)
\end{eqnarray}
and
\begin{eqnarray}
\bar{A}^{+-} & \equiv & \langle D^+D^- |{\cal H}_{\rm eff}| \bar{B}^0_d\rangle \; = \;
\frac{1}{2} \left ( \bar{A}_1 ~ + ~ \bar{A}_0 \right ) \; , \nonumber \\
\bar{A}^{00} & \equiv & \langle D^0\bar{D}^0 |{\cal H}_{\rm eff}| \bar{B}^0_d\rangle \; = \;
\frac{1}{2} \left ( \bar{A}_1 ~ - ~ \bar{A}_0 \right ) \; , \nonumber \\
\bar{A}^{-0} & \equiv & \langle D^-D^0 |{\cal H}_{\rm eff}| B^-_u\rangle 
\; = \; \bar{A}_1 \; .
%		(4.2)
\end{eqnarray}
The isospin relations (4.1) and (4.2) form two triangles in the complex plane:
\begin{eqnarray}
A^{+-} ~ + ~ A^{00} & = & A^{+0} \; , \nonumber \\
\bar{A}^{+-} ~ + ~ \bar{A}^{00} & = & \bar{A}^{-0} \; .
%		(4.3)
\end{eqnarray}
One is able to determine the relative size and phase difference of 
isospin amplitudes $A_1$ ($\bar{A}_1$) and $A_0$ ($\bar{A}_0$) from the
above triangular relations. Denoting
\begin{equation}
\frac{A_0}{A_1} \; \equiv \; z e^{{\rm i}\theta} \; , ~~~~~~~~
\frac{\bar{A}_0}{\bar{A}_1} \; \equiv \; \bar{z} e^{{\rm i} \bar{\theta}} \; ,
%		(4.4)
\end{equation}
then we obtain
\begin{eqnarray}
z & = & \sqrt{\frac{2 \displaystyle \left ( |A^{+-}|^2 + |A^{00}|^2 \right )}
{|A^{+0}|^2} ~ - ~ 1} \; , \nonumber \\
\theta & = & \arccos \left ( \frac{|A^{+-}|^2 - |A^{00}|^2}
{z ~ |A^{+0}|^2} \right ) \; ; 
%		(4.5)
\end{eqnarray}
and
\begin{eqnarray}
\bar{z} & = & \sqrt{\frac{2 \left ( |\bar{A}^{+-}|^2 + |\bar{A}^{00}|^2 \right )}
{|\bar{A}^{-0}|^2} ~ - ~ 1} \; , \nonumber \\
\bar{\theta} & = & \arccos \left ( \frac{|\bar{A}^{+-}|^2 - |\bar{A}^{00}|^2}
{\bar{z} ~ |\bar{A}^{-0}|^2} \right ) \; .
%		(4.6)
\end{eqnarray}
If $z=1$ and $\theta =0$, e.g., we find that $|A^{00}|=0$ (i.e.,
the decay mode $B^0_d\rightarrow D^0\bar{D}^0$ is forbidden).
Note that $\theta$ ($\bar{\theta}$) is in general a mixture of the weak and
strong phase shifts, since both $A_0$ ($\bar{A}_0$) and $A_1$ ($\bar{A}_1$)
may contain the tree-level and penguin contributions. 

It is worth pointing out that the same isospin relations hold for the decay
modes $B\rightarrow D\bar{D}^*$ and $B\rightarrow D^*\bar{D}$. Of course,
the isospin parameters $z$ ($\bar{z}$) and $\theta$ ($\bar{\theta}$) in
$B\rightarrow D\bar{D}$, $D\bar{D}^*$ and $D^*\bar{D}$ may be different 
from one another due to their different final-state interactions. 
As for $B\rightarrow D^*\bar{D}^*$,
the same isospin relations hold separately for the decay amplitudes
with helicity $\lambda=-1$, $0$, or $+1$.

The quantities $|A^{+0}|$ and $|\bar{A}^{-0}|$ are obtainable from
the time-independent measurements of 
decay rates of $B^+_u\rightarrow D^+\bar{D}^0$ and
$B^-_u\rightarrow D^-D^0$. A determination of
$|A^{+-}|$ ($|A^{00}|$) and $|\bar{A}^{+-}|$ ($|\bar{A}^{00}|$) is possible
through the time-integrated measurements of $B^0_d$ vs $\bar{B}^0_d\rightarrow 
D^+D^-$ ($D^0\bar{D}^0$) 
on the $\Upsilon (4S)$ resonance, where the produced two $B_d$ mesons are
in a coherent state (with odd charge-conjugation parity) until one of them 
decays. In practice, one can use 
the semileptonic transition of one $B_d$ meson to tag the flavor 
of the other meson decaying to $D^+D^-$ or $D^0\bar{D}^0$. 
To probe the $CP$ asymmetry induced by the interplay of direct decay
and $B^0_d$-$\bar{B}^0_d$ mixing in $B_d\rightarrow D\bar{D}$, the time-dependent
measurements are necessary on the $\Upsilon (4S)$ resonance at asymmetric
$B$ factories. In such an experimental scenario, the joint decay rates can be given as
follows \cite{XingB96,SandaXing}:
\begin{eqnarray}
{\cal R}(l^{\pm}X^{\mp}, D^+D^-; t) & \propto & |A_l|^2 e^{-\Gamma |t|} \left [
\frac{|A^{+-}|^2 + |\bar{A}^{+-}|^2}{2} ~ \mp ~ \frac{|A^{+-}|^2 - |\bar{A}
^{+-}|^2}{2} \cos (x_d \Gamma t) \right . \nonumber \\
&  & \left . \pm ~ |A^{+-}|^2 ~ {\rm Im} \left ( \frac{q^{~}_B}{p^{~}_B}
\frac{\bar{A}^{+-}}{A^{+-}} \right ) \sin (x_d \Gamma t) \right ] \; 
%		(4.7)
\end{eqnarray}
and
\begin{eqnarray}
{\cal R}(l^{\pm}X^{\mp}, D^0\bar{D}^0; t) & \propto & |A_l|^2 e^{-\Gamma |t|} \left [
\frac{|A^{00}|^2 + |\bar{A}^{00}|^2}{2} ~ \mp ~ \frac{|A^{00}|^2 - |\bar{A}
^{00}|^2}{2} \cos (x_d \Gamma t) \right . \nonumber \\
&  & \left . \pm ~ |A^{00}|^2 ~ {\rm Im} \left ( \frac{q^{~}_B}{p^{~}_B}
\frac{\bar{A}^{00}}{A^{00}} \right ) \sin (x_d \Gamma t) \right ] \; , 
%		(4.8)
\end{eqnarray}
where $t$ is the proper time difference between the semileptonic and nonleptonic
decays
\footnote{Note that the proper time sum of the semileptonic and nonleptonic 
decays has been integrated out, since it will not be measured at any $B$-meson factory.}.
Denoting
\begin{equation}
\phi_{DD} \; \equiv \; \frac{1}{2} \arg \left (\frac{q^{~}_B}{p^{~}_B} ~
\frac{\bar{A}_1}{A_1} \right ) \; ,
%		(4.9)
\end{equation}
we express coefficients of the $\sin (x_d \Gamma t)$ term in (4.7) and (4.8) 
in terms of isospin parameters:
\begin{eqnarray}
{\rm Im} \left ( \frac{q^{~}_B}{p^{~}_B} \frac{\bar{A}^{+-}}{A^{+-}} \right ) & = &
\frac{|A^{+0} \bar{A}^{-0}|}{4 |A^{+-}|^2} \left [ \sin \left (2\phi_{DD} \right ) ~ - ~
z \sin \left (\theta - 2\phi_{DD} \right ) \right . \nonumber \\
&  & \left . + ~ \bar{z} \sin \left (\bar{\theta} + 2\phi_{DD} \right ) ~ + ~ z \bar{z} 
\sin \left (\bar{\theta} - \theta + 2\phi_{DD} \right ) \right ] \; 
%		(4.10)
\end{eqnarray}
and
\begin{eqnarray}
{\rm Im} \left ( \frac{q^{~}_B}{p^{~}_B} \frac{\bar{A}^{00}}{A^{00}} \right ) & = &
\frac{|A^{+0} \bar{A}^{-0}|}{4 |A^{00}|^2} \left [ \sin \left (2\phi_{DD} \right ) ~ + ~
z \sin \left (\theta - 2\phi_{DD} \right ) \right . \nonumber \\
&  & \left . - ~ \bar{z} \sin \left (\bar{\theta} + 2\phi_{DD} \right ) ~ + ~ z \bar{z} 
\sin \left (\bar{\theta} - \theta + 2\phi_{DD} \right ) \right ] \; .
%		(4.11)
\end{eqnarray}
All the quantities on the right-hand side of (4.10) or (4.11), except $\phi_{DD}$, 
can be determined through the time-independent
measurements of $B\rightarrow D\bar{D}$ on the $\Upsilon (4S)$ resonance. Thus 
measuring the $CP$-violating observable on the left-hand side of (4.10) or (4.11)
will allow a model-independent extraction of $\phi_{DD}$.

Two remarks about the results obtained above are in order:

(1) If the tree-level quark transition
$\bar{b}\rightarrow (c\bar{c})\bar{d}$ is assumed to dominate the decay amplitude of 
$B^+_u\rightarrow D^+\bar{D}^0$, i.e., $\bar{A}_1/A_1 \approx (V_{cb}V^*_{cd})/(V^*_{cb}V_{cd})$,
then we get $\phi_{DD}\approx \beta$ as a pure weak
phase in the standard model. 
In general, $\phi_{DD}$ should be a mixture of both weak and strong phases
due to the penguin effects \cite{SandaXing}. 
We expect that a comparison of $\phi_{\psi K}$ 
(extracted from $B_d\rightarrow \psi K_S$ or $\psi K_L$) with
$\phi_{DD}$ (extracted from $B_d\rightarrow D^+D^-$ or $B_d\rightarrow D^0\bar{D}^0$) 
would constrain the penguin-induced phase information in $B\rightarrow D\bar{D}$.

(2) A special but interesting case is $z = \bar{z} =1$. It can be obtained if 
the decay modes $B\rightarrow D\bar{D}$ occur dominantly through the tree-level
subprocess $b\rightarrow (c\bar{c}) d$ or $\bar{b}\rightarrow (c\bar{c}) \bar{d}$.
In this case, $A_0$ ($\bar{A}_0$) and $A_1$ ($\bar{A}_1$) have a common KM factor;
thus $\theta$ ($\bar{\theta}$) is a 
pure strong phase shift. This will lead, for arbitrary values of $\theta$ and 
$\bar{\theta}$, to the relations
\begin{eqnarray}
|A^{+-}|^2 ~ + ~ |A^{00}|^2 & = & |A^{+0}|^2 \; , \nonumber \\
|\bar{A}^{+-}|^2 ~ + ~ |\bar{A}^{00}|^2 & = & |\bar{A}^{-0}|^2 \; ;
%		(4.12)
\end{eqnarray}
i.e., the two isospin triangles in (4.3) become right-angled triangles.
If $\theta = \bar{\theta}$ is further assumed, we obtain
\begin{eqnarray}
{\rm Im} \left ( \frac{q^{~}_B}{p^{~}_B} \frac{\bar{A}^{+-}}{A^{+-}} \right ) & = &
\frac{|A^{+0} \bar{A}^{-0}|}{|A^{+-}|^2} \sin \left (2\phi_{DD} \right ) 
\cos^2 \frac{\theta}{2} \; , \nonumber \\
{\rm Im} \left ( \frac{q^{~}_B}{p^{~}_B} \frac{\bar{A}^{00}}{A^{00}} \right ) & = &
\frac{|A^{+0} \bar{A}^{-0}|}{|A^{00}|^2} \sin \left (2\phi_{DD} \right ) 
\sin^2 \frac{\theta}{2} \; .
%		(4.13)
\end{eqnarray}
One can see that these two $CP$-violating quantities have the quasi-seesaw 
dependence on the isospin phase shift $\theta$. The magnitude of $\sin (2\phi_{DD})$
turns out to be
\begin{equation}
\sin(2\phi_{DD}) \; = \; - \frac{1}{|A^{+0}\bar{A}^{-0}|} \left [
|A^{+-}|^2 {\rm Im}\left (\frac{q^{~}_B}{p^{~}_B} \frac{\bar{A}^{+-}}{A^{+-}} \right )
~ + ~ |A^{00}|^2 {\rm Im} \left (\frac{q^{~}_B}{p^{~}_B} \frac{\bar{A}^{00}}{A^{00}} \right )
\right ] \; ,
%		(4.14)
\end{equation}
apparently independent of $\theta$.

%======================================================================
\section{Probing $\sin (2\beta)$ in $B_d\rightarrow (f_{CP})_D + ( \pi^0, ~ \rho^0,
~ a^0_1)$}
\setcounter{equation}{0}

The third type of $B_d$ decays for the extraction of $\sin (2\beta)$ is expected to be
\cite{DunietzSnyder91}
\be
B_d \; \longrightarrow \; (f_{CP})_D ~ + ~ (\pi^0, ~ \rho^0, ~ a^0_1, ~ {\rm etc}) \; ,
%		(5.1)
\ee
where the $CP$ eigenstate $f_{CP}$ may come from either $D^0$ or $\bar{D}^0$ in the 
neglect of non-trivial $D^0$-$\bar{D}^0$ mixing effects \cite{XingBD96,XingD97}. 
Such transitions occur only through 
the tree-level quark diagrams, as illustrated by Fig. 3. We observe that the graph 
amplitudes of Fig. 3(a) are doubly KM-suppressed with respect to those of Fig. 3(b),
and the ratio of their KM factors is $|V_{cd}/V_{ud}|\cdot |V_{ub}/V_{cb}| \approx
2\%$ \cite{PDG} in the standard model. Therefore, the contribution from Fig. 3(a) can be
safely neglected in discussing indirect $CP$ violation induced by the interplay of
decay and $B^0_d$-$\bar{B}^0_d$ mixing
\footnote{Direct $CP$ violation may appear due to interference between the graph
amplitudes of Fig. 3(a) and Fig. 3(b). These two amplitudes have different isospin structures,
hence a strong phase shift between them (denoted by $\Delta$) is possible as the necessary
ingredient of a direct $CP$ asymmetry (proportional to $\sin\Delta$). However, there is
no way to evaluate this strong phase theoretically. Even if $|\sin\Delta|\sim 1$, the
$CP$ asymmetry is at most of the percent level because of the KM suppression for the
graph amplitudes in Fig. 3(a).}.

We remark the assumption that possible effects induced by $D^0$-$\bar{D}^0$ mixing 
are negligible in the $B_d$ decay modes under consideration. The latest constraint
on the $D^0$-$\bar{D}^0$ mixing rate is $r^{~}_D < 0.5\%$ \cite{Dmixing,Golowich96}, which can
be safely neglected for our present purpose. In case that the mixing phase $q^{~}_D/p^{~}_D$
were nonvanishing, it would give rise to measurable $CP$ asymmetries in some neutral
$D$-meson decays to $CP$ eigenstates $f_{CP}$ (such as $f_{CP} = \pi^+\pi^-$, $K^+K^-$
and $K_S \pi^0$) \cite{XingD97,BigiSanda86}. 
The current limits on the asymmetries between the decay rates of
$D^0\rightarrow f_{CP}$ and $\bar{D}^0\rightarrow f_{CP}$ show no $CP$ violation at
the percent level \cite{DCP}. If we further assume that the penguin amplitude of $D^0\rightarrow
f_{CP}$ is not enhanced by possible new physics \cite{DunietzSnyder91}, 
i.e., $D^0\rightarrow f_{CP}$ occurs
dominantly through the tree-level quark diagrams with a single KM factor, 
then the overall amplitudes of $B^0_d\rightarrow (f_{CP})_{\bar{D}^0} + \pi^0$ and
$\bar{B}^0_d\rightarrow (f_{CP})_{D^0} + \pi^0$ can be written as folllows:
\ba
\langle (\pi^+\pi^-)_{\bar{D}^0} ~ \pi^0 |{\cal H}_{\rm eff}| B^0_d\rangle & = & 
\left (V_{cb}^* V_{ud} \right ) \left (V_{cd}V_{ud}^* \right )  A_{D\pi} ~ 
A_{\pi\pi} \; , \nonumber \\
\langle (K^+K^-)_{\bar{D}^0} ~ \pi^0 |{\cal H}_{\rm eff}| B^0_d\rangle & = & 
\left (V_{cb}^* V_{ud} \right ) \left (V_{cs}V_{us}^* \right )  A_{D\pi} ~ 
A_{KK} \; , \nonumber \\
\langle (K_S\pi^0)_{\bar{D}^0} ~ \pi^0 |{\cal H}_{\rm eff}| B^0_d\rangle & = & 
\left (V_{cb}^* V_{ud} \right ) \left (V_{cs}V_{ud}^* ~ p^*_K \right ) 
A_{D\pi} ~ A_{K\pi} \; , 
%		(5.2)
\ea
and
\ba
\langle (\pi^+\pi^-)_{D^0} ~ \pi^0 |{\cal H}_{\rm eff}| \bar{B}^0_d\rangle & = & 
- \left (V_{cb} V_{ud}^* \right ) \left (V_{cd}^* V_{ud} \right )  A_{D\pi} ~ 
A_{\pi\pi} \; , \nonumber \\
\langle (K^+K^-)_{D^0} ~ \pi^0 |{\cal H}_{\rm eff}| \bar{B}^0_d\rangle & = & 
- \left (V_{cb} V_{ud}^* \right ) \left (V_{cs}^* V_{us} \right )  A_{D\pi} ~ 
A_{KK} \; , \nonumber \\
\langle (K_S\pi^0)_{D^0} ~ \pi^0 |{\cal H}_{\rm eff}| \bar{B}^0_d\rangle & = & 
+ \left (V_{cb} V_{ud}^* \right ) \left (V_{cs}^* V_{ud} ~ q^*_K \right ) 
A_{D\pi} ~ A_{K\pi} \; .
%		(5.3)
\ea
Here $A_{D\pi}$, $A_{\pi\pi}$, $A_{KK}$ and $A_{K\pi}$ denote the hadronic 
matrix elements containing strong interaction phases; $p^{~}_K$ and $q^{~}_K$ 
are the $K^0$-$\bar{K}^0$ mixing parameters; and the ``$\pm$'' sign arises from
the $CP$-even or $CP$-odd final state. Let us define three phase observables:
\ba
\phi_{\pi\pi} & \equiv & \frac{1}{2} \arg \left (\frac{q^{~}_B}{p^{~}_B} ~
\frac{V_{cb} V^*_{cd}}{V_{cb}^* V_{cd}} \right ) \; , \nonumber \\
\phi_{KK} & \equiv & \frac{1}{2} \arg \left ( \frac{q^{~}_B}{p^{~}_B} ~
\frac{V_{cb} V^*_{ud}}{V_{cb}^* V_{ud}} ~ \frac{V^*_{cs} V_{us}}{V_{cs} V^*_{us}}
\right ) \; , \nonumber \\
\phi_{K\pi} & \equiv & \frac{1}{2} \arg \left (\frac{q^{~}_B}{p^{~}_B} ~
\frac{q^*_K}{p^*_K} ~ \frac{V_{cb} V^*_{cs}}{V_{cb}^* V_{cs}} \right ) \; .
%		(5.4)
\ea
At an asymmetric $B$ factory running on the $\Upsilon (4S)$ resonance, one 
can measure the following joint decay rates to extract $\phi_{\pi\pi}$,
$\phi_{KK}$ and $\phi_{K\pi}$:
\ba
{\cal R} (l^{\pm}X^{\mp}, (\pi^+\pi^-)_D ~ \pi^0; t) & \propto &
|A_l|^2  e^{-\Gamma |t|} \left [ 1 ~ \mp ~
\sin (2\phi_{\pi\pi}) \cdot \sin (x_d \Gamma t) \right ] \; , \nonumber \\
{\cal R} (l^{\pm}X^{\mp}, (K^+K^-)_D ~ \pi^0; t) & \propto &
|A_l|^2  e^{-\Gamma |t|} \left [ 1 ~ \mp ~
\sin (2\phi_{KK}) \cdot \sin (x_d \Gamma t) \right ] \; , \nonumber \\
{\cal R} (l^{\pm}X^{\mp}, (K_S\pi^0)_D ~ \pi^0; t) & \propto &
|A_l|^2  e^{-\Gamma |t|} \left [ 1 ~ \pm ~
\sin (2\phi_{K\pi}) \cdot \sin (x_d \Gamma t) \right ] \; ,
%		(5.5)
\ea
where the semileptonic modes ($l^{\pm}X^{\mp}$) serve for the flavor tagging
of $B_d$ mesons, and $t$ is the proper time difference between the semileptonic
and nonleptonic decays. Of course, these decay rates have different normalization
factors.
 
The feasibility to measure the joint decay modes in (5.5) depends crucially upon
the branching ratio of $B^0_d\rightarrow \bar{D}^0\pi^0$ and that of $\bar{D}^0
\rightarrow f_{CP}$. The latter has been determined in experiments of charm physics \cite{PDG}.
Current data only yield the upper bound ${\cal B}(B_d^0\rightarrow \bar{D}^0\pi^0)
< 4.8\times 10^{-4}$ \cite{PDG}. The lower bound of ${\cal B}(B^0_d\rightarrow 
\bar{D}^0\pi^0)$ is obtainable from an isospin analysis of $B^0_d\rightarrow
\bar{D}^0\pi^0$, $B^0_d\rightarrow D^-\pi^+$ and $B^+_u\rightarrow \bar{D}^0\pi^+$.
One can easily find 
\ba
\langle D^-\pi^+ |{\cal H}_{\rm eff}|B^0_d\rangle & = & A_{3/2} ~ + ~ \sqrt{2}
A_{1/2} \; , \nonumber \\
\langle \bar{D}^0 \pi^0 |{\cal H}_{\rm eff}|B^0_d\rangle & = & \sqrt{2} A_{3/2}
~ - ~ A_{1/2} \; , \nonumber \\
\langle \bar{D}^0 \pi^+ |{\cal H}_{\rm eff}|B^+_u\rangle & = & 3 A_{3/2} \; ,
%		(5.6)
\ea
where $A_{3/2}$ and $A_{1/2}$ stand respectively for the $I=3/2$ and $I=1/2$ isospin
amplitudes with the common KM factor $V^*_{cb}V_{ud}$. Neglecting tiny isospin-violating
effects induced by the mass differences $m^{~}_{D^0}-m^{~}_{D^+}$ and $m^{~}_{\pi^0}
-m^{~}_{\pi^+}$ as well as the life time difference $\tau^{~}_{B_d}-\tau^{~}_{B_u}$,
we get from (5.6) that \cite{XingIso95}
\be
{\cal B} (B^0_d\rightarrow \bar{D}^0\pi^0) \; \geq \; \frac{1}{2} \left [
1 ~ - ~ \sqrt{\frac{{\cal B}(B^0_d\rightarrow D^-\pi^+)}{{\cal B}(B^+_u\rightarrow
\bar{D}^0\pi^+)}} \right ]^2 {\cal B}(B^+_u\rightarrow \bar{D}^0\pi^+) \; . 
%		(5.7)
\ee
Since ${\cal B}(B^0_d\rightarrow D^-\pi^+) = (3.0\pm 0.4)\times 10^{-3}$ and
${\cal B}(B^+_u\rightarrow \bar{D}^0\pi^+)=(5.3\pm 0.5)\times 10^{-3}$ have been
measured \cite{PDG}, we are able to obtain the lower bound of ${\cal B}(B^0_d\rightarrow
\bar{D}^0\pi^0)$ model-independently, as numerically illustrated in Fig. 4.
This result implies that the decay mode $B^0_d\rightarrow \bar{D}^0\pi^0$ should
be detected soon.

In practice, it is necessary to sum over all possible decay modes of the same
nature as $B^0_d\rightarrow \bar{D}^0\pi^0$, such as $B^0_d\rightarrow \bar{D}^0
\rho^0$ and $\bar{D}^0a^0_1$. These transitions are governed by the same weak
interactions, thus their branching ratios are expected to be of the same order
\cite{XingIso95,Yamamoto94}.
It is also a good idea to sum over all possible $\bar{D}^0\rightarrow f_{CP}$
decays of the same nature, e.g., $f_{CP} = K_S \pi^0$, $K_S \rho^0$, $K_S \omega$,
etc \cite{DunietzSnyder91}. For a careful classification of $CP$ parities in the final states of
$B_d\rightarrow (f_{CP})_D + (\pi^0, \rho^0, ~{\rm etc})$, we refer the reader
to Ref. \cite{DunietzSnyder91}.

%=============================================================================
\section{Concluding remarks}
\setcounter{equation}{0}

We have discussed three different possibilities to determine the $CP$-violating
quantity $\sin (2\beta)$ in the first-round experiments of a $B$-meson factory.
They should be supplementary to the gold-plated approach, where $\sin (2\beta)$
is related to the $CP$ asymmetry in $B_d\rightarrow \psi K_S$ or 
$B_d\rightarrow \psi K_L$ within the standard model. In addition, it has been
pointed out that the magnitude of $\sin (2\beta)$ can be well constrained 
from a variety of quark mass ans$\rm\ddot{a}$tze. Some necessary remarks about the
results obtained above are in order.

(a) The uncertainties associated with the approximate relations (2.3), (2.4)
and (2.7) are expected to be less than $10\%$, as a consequence of the significant
hierarchy of quark mass values. This accuracy should be good enough to justify
or rule out the relevant quark mass ans$\rm\ddot{a}$tze, if $\sin (2\beta)$ 
can be measured to the similar extent of precision.

(b) The $CP$-violating signal ${\cal S}_{CP}$ in (3.10) is worth being pursued
experimentally. If there exist some difficulties in detecting it within the
first-round experiments of a $B$ factory, further efforts should be made in
the second-round experiments. Some other $CP$-forbidden channels of $B_d\bar{B}_d$
decays on the $\Upsilon (4S)$ resonance are also interesting for the study of
$CP$ violation.

(c) Within the standard model, we have 
$\phi_{\pi\pi} \approx \phi_{KK} \approx \phi_{K\pi} \approx \beta$ to an
excellent degree of accuracy. The deviation of $\phi_{DD}$ from $\phi_{\psi K}
= \phi_{K\pi}$ might not be negligibly small, provided the penguin effects in
$B_d\rightarrow D\bar{D}$ were not as small as we naively expected.
New physics in $K^0$-$\bar{K}^0$ mixing could give rise to an observable 
difference between $\phi_{\pi\pi}$ ($\phi_{KK}$) and $\phi_{K\pi}$ or between
$\phi_{\psi K}$ and $\phi_{DD}$. New physics in $B^0_d$-$\bar{B}^0_d$ mixing
would affect all the five $CP$-violating phases under discussion, but could not
be isolated from the proposed measurements.

Of course, we have assumed unitarity of the $3\times3$ KM matrix in the above
discussions. Some of our results are indeed independent of this assumption.
New physics, which can violate the KM unitarity and in turn affect $CP$
asymmetries of some $B_d$ decays, has been classified in 
Ref. \cite{Grossman}. 

\vspace{0.5cm}

\begin{flushleft}
{\Large\bf Acknowledgments}
\end{flushleft}

It is my pleasure to thank the chairman of this workshop, Y. Koide, for
his invitation and hospitality. I am grateful to 
H. Fritzsch and A.I. Sanda, who shared physical ideas with me in
Refs. \cite{FritzschXing95} and \cite{SandaXing} respectively, for
some useful discussions. A valuable conversation with Y. Iwasaki on the
KEK $B$-factory program is also acknowledged. This work was supported by
the Japan Society for the Promotion of Science.

\end{document}